\documentclass[10pt,a4paper]{article}
\usepackage[utf8]{inputenc}
\usepackage{amsmath}

\usepackage{amsthm}
\usepackage{amsfonts}
\usepackage{amssymb}
\usepackage{graphicx}
\usepackage{cleveref}
\usepackage{caption}
\usepackage{subfigure}
\usepackage{epstopdf}
\usepackage{todonotes}
\usepackage{siunitx}

\theoremstyle{definition}

\begin{document}
	\title{\LARGE \bf YatSim: an Open-Source Simulator For Testing Consensus-based Control Strategies in Urban Traffic Networks}
	\author{Alexander Martin Dethof
		\thanks{A. M. Dethof is with the Control Systems Group - Technische Universität Berlin, Germany. alexander.m.dethof@campus.tu-berlin.de}
		\and Fabio Molinari
		\thanks{F. Molinari is with the Control Systems Group - Technische Universität Berlin, Germany. molinari@control.tu-berlin.de. 
			This work was funded by the German Research Foundation (DFG) within their priority programme SPP 1914 "Cyber-Physical 
			Networking (CPN)", RA516/12-1.
		}
	}
	\maketitle
	
	\section{Introduction}
	
		Autonomous driving promises a variety of useful advantages. Its realization targets to be more likely than it was so far \cite{Schoettle.2014}. Current statistics predict for the year 2050 nearly each vehicle to ride fully-autonomously \cite{IHSAutomotive.2014}. At present, currently developed autonomous assistance systems \cite{Sugeno.1984} \cite{Yi.2001} \cite{Naranjo.2008} aim to aid within the current traffic environment. However, this environment is adopted on human demands. Instead, a vehicle-adopted infrastructure, would leverage the efficiency of fully-autonomous driving. Currently, road intersections are highly inefficient components. Despite of modern traffic signalling approaches, they are not able to suitably react on sudden condition changes.
		
		Thus, the present work concentrates on the CBAA-M algorithm proposed by Molinari et al. \cite{Molinari.2018}, which enables multi-agent cooperation within autonomous traffic to safely and efficiently pass intersections. In detail, it supports autonomous vehicles to independently achieve consensus with the surrounding agents about the passing order for a given intersection. Based on the Model Predictive Control (MPC) approach of Katriniok et al. \cite{Katriniok.2017}, each agent is able to adapt its individual movements in order to safely and efficiently pass the intersection in accordance with the corresponding passing order. Upon this method, we developed an open-source software framework, which is able to simulate and validate CBAA-M the \cite{Molinari.2018} within a realistic urban scenario. Due to the variety of existing traffic simulator applications, we named our solution \emph{yatSim}, i.e. `yet another traffic simulator'.
		
		This paper aims to present the work on yatSim in detail and outline issues for future integrations into real systems. We thus provide in Section \ref{sec:rel-work} a brief overview about related simulator approaches and indicate their differences among our application. Upon this, Section \ref{sec:cbaa-m} repeats the basics of the CBAA-M, whereas Section \ref{sec:concept} outlines the simulation concept and indicates issues of the development process. The paper concludes with a presentation of a sample simulation in Section \ref{sec:simulation} and yields a final review in Section \ref{sec:conclusion}.
	
		\subsection{Terms}
			Considering traffic simulation, two implementation possibilities apply: \emph{macroscopic} and \emph{microscopic} simulation \cite{Pursula.1999}. Macroscopic simulation concentrates on the simulation of flow-based models. Whereas microscopic simulation computes the behaviour of each vehicle individually. This is especially relevant to test new approaches in small areas, whereas macroscopic approaches investigate the behavioural impact for broader areas, such as complete cities. With a high amount of vehicles, macroscopic results can be also gathered from microscopic simulations. However, with increasing traffic scenarios, higher computational efforts are necessary. Hence, flow-based models are preferred in those situations.

	\section{Related Work}\label{sec:rel-work}
		Kotusevski et al. \cite{Kotusevski.2009} provides a detailed overview and evaluation about existing traffic simulators: Simulation Of Urban Mobility (SUMO), Quadstone Paramics Modeller, Treiber's Microsimulation of Road Traffic, Aimsun, Trafficware SimTraffic and CORSIM TRAFVU. According to the licensing issue, only two of these six packages are available as open source projects: SUMO \cite{Behrisch.2011} and Treiber’s Microsimulation \cite{Treiber.2009}. From these two, only SUMO promises a useful extendibility. Furthermore, it is also the only one, which is able to import real urban maps. According to \cite{Krajzewicz.2012}, SUMO is able to handle different traffic control policies, but still based on common, hence inefficient, traffic lights.
		
		A possible integration of CBAA-M \cite{Molinari.2018} into SUMO would thus require a deeper initial 	training. To the best of our knowledge, CBAA-M is not validated so far within urban scenarios. Hence, implementing a non-validated algorithm into a complex software as SUMO could cause additional development issues. As we aim to investigate CBAA-M's capability within realistic situations, we decided to start a step earlier and present with yatSim a new simulator. However, after a successful validation, an integration into well-developed approaches as SUMO, is feasible and thus highly welcomed.
	
	\section{CBAA-M}\label{sec:cbaa-m}
		As \cite{Molinari.2018} already states a detailed insight into the CBAA-M, we hereby deliver a brief abstract of the algorithm itself to explain yatSim's principal logic. In particular the algorithm is derived from the robotics-based CBAA-approach of Choi et al. \cite{Choi.2009}. It consists of two main phases: an auction- and a consensus-based moment. The auction moment builds upon a market-based selection strategy, where each vehicle $i \in S$ computes a bid $c_i(k)$ - which bases on a linear combination of the distance to a common `collision point', i.e. a shared coordinate on the future path of two agents, and the current speed - for a sampled time instant $k \in \mathbb{N}$. $S$ hereby indicates the set of all vehicles, which are currently known to the system. In a subsequent auction and consensus process, the vehicles decentrally sort the bids of each agent in descending order and thus obtain an ordered priority list, which indicates the order the vehicle are allowed to pass the dedicated intersection.
		
		\begin{align}
			\mathbf{x}_i(k + 1) &= \mathbf{x}_i(k) + T_s \cdot \mathbf{v}_i(k) \label{eq:vehicle-dynamics:1} \\
			\mathbf{v}_i(k + 1) &= \mathbf{v}_i(k) + T_s \cdot \mathbf{a}_i(k) \label{eq:vehicle-dynamics:2}
		\end{align}
		
		An on-board mounted MPC based on the Katrinok et al. \cite{Katriniok.2017} computes - based on the dynamics model in Equation \ref{eq:vehicle-dynamics:1} and \ref{eq:vehicle-dynamics:2} - the acceleration $\mathbf{a}_i$, agent $i$ requires in order to assure the passing order and avoid rear-end-collisions with low impact on the overall traffic flow. $\mathbf{x}_i$ hereby indicates $i$'s position and $\mathbf{v}_i$ the corresponding speed. $T_s$ states the sample time passed between the two time steps $\left(k + 1\right)$ and $k$.
	
	\section{Concept}\label{sec:concept}
	
		Based on the above described approach, we developed the yatSim application in order to validate the algorithm within realistic urban scenarios.

		\subsection{Application Description}
			As initial development parameter, we decided to chose \emph{Python} as implementation language, since it is broadly known within the scientific community, portable between different operating systems and free. Furthermore, for Python exists the \emph{CVXPY}\footnote{http://www.cvxpy.org/en/latest} module, which enables enhanced integration of MPC-based solution design. As surrounding graphical framework, we chose the \emph{Kivy}\footnote{http://kivy.org} framework, which features a clear separation between graphical representation and internal application logic. Futhermore, it automatically creates appropriate event handlers, which enables a simple, fast and sustainable development process.
			
			At the beginning of the development process, we specified different requirements among the final application. As first aim, we propose \emph{advanced usability}, i.e. inexperienced users should be able to quickly and easily generate complex simulation scenarios without a long initial training. Furthermore, we aimed to model \emph{realistic scenarios} with multiple intersections, randomized vehicle flows and paths of different turning requirements. In this context, we further aimed to implement a microscopic simulation, since \cite{Molinari.2018} delivers a vehicle-based description of the CBAA-M. Upon that, we demanded a multi-threaded environment to feature \emph{realistic microscopic simulations}, where each thread represents a corresponding vehicle. Finally, we claimed \emph{repeatability and reproducibility}, which demands the application to save and reload previously generated simulation maps.
			
\begin{figure}
				\includegraphics[width=\columnwidth]{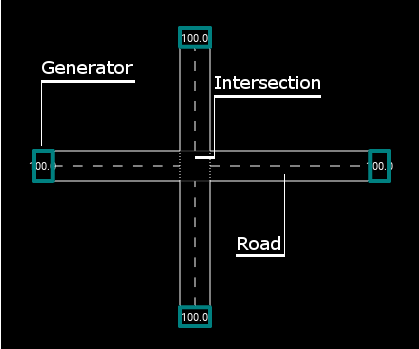}
				\caption{The traffic elements of yatSim: \emph{generators} can introduce new \emph{vehicles} into the environment, which are able to move on the surrounding \emph{roads} and \emph{intersections}}
				\label{fig:elements}
				\vspace{-13pt}
			\end{figure}

			Initially, yatSim supports the following main components as in the Figure \ref{fig:elements}: two types of roads - a horizontally- and a vertically-oriented one, an intersection component to enables both road types to cross each other and a generator component at the end of each road to introduce new vehicles to the system. With these components the user is able to quickly generate in a first approach `Manhattan'-like orthogonal traffic infrastructures. More complex structures are not implemented so far, but are integrated within future releases.
			
\begin{figure}
				\includegraphics[width=\columnwidth]{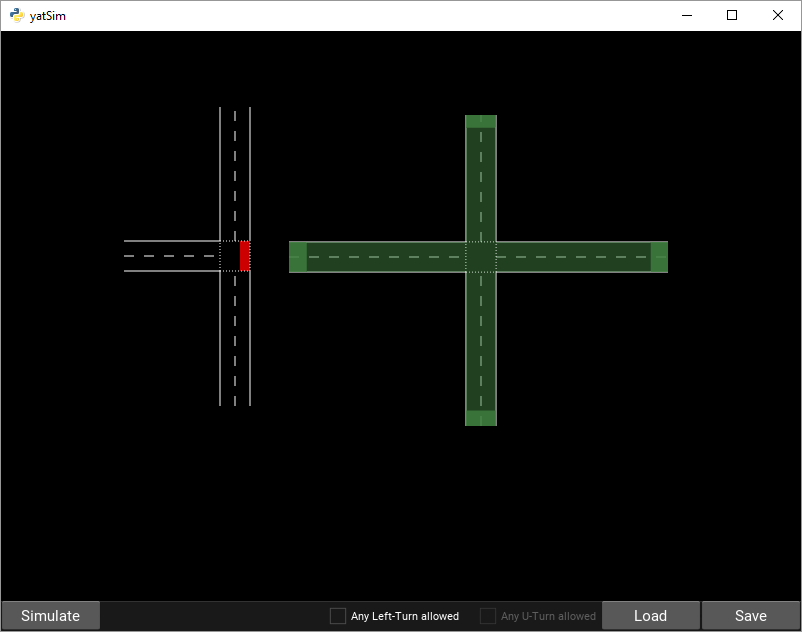}
				\caption{A broken traffic infrastructure: The two intersection were formerly connected in one scenario. By moving the highlighted (green) intersection, the network broke, which is identified by the red corner in the left intersection - yatSim can repair the scenario by creating a new road at the broken intersection}
				\label{fig:broken-traffic-infrastructure}
				\vspace{-13pt}
			\end{figure}

			To assure simple user interaction and hence an advanced usability, the user can design traffic scenarios with a single click into the application window and a subsequent mouse-move action. As far as a horizontal and a vertical road crosses each other, the application automatically replaces this point with a corresponding intersection component. Pressing the key `G' during a click places a generator at the end of the road located beneath the cursor. As a further feature, the components are manipulatable by moving or resizing action. If the logic of the traffic structure breaks, as in the Figure \ref{fig:broken-traffic-infrastructure}, due to such a manipulative action, yatSim automatically detects the initial trigger, highlights it and delivers a solution to repair.
			
		\subsection{Simulation Process}
			The yatSim engine performs the traffic simulation within a time-discrete process. Therefore it uses a clock to trigger each millisecond a `tick'-event. Since the computational complexity of one time-frame varies with the scenario's structure and the amount of vehicles in the system, the execution of the actual simulation tick occurs within a dedicated parallel thread. Before a new `tick'-event starts, the simulator checks, if the former `tick' has already completed. Only in this case, the simulator triggers the next `tick'-thread execution.
			
			During a simulation ‘tick’, yatSim iterates through the scenario's generators and vehicles to forward the ‘tick’-trigger. Each component-based `tick' is then executed within a further parallel thread. To ensure newly generated vehicles to be simulated with appropriate starting values, we apply the rule that already existing vehicles are triggered before the actual generators. This prevents a vehicle to immediately move within the same simulation cycle it was created in.
			
			The vehicle generation bases on a Monte-Carlo principle: During scenario setup, the user specifies for each generator a probability $p$ that a vehicle is generated within a dedicated simulation cycle. The generator computes during each execution a random decimal number $r \in \left[0, 1\right]$. If $1 - r \geq p$, the generator introduces a new vehicle into the system. In order to choose a feasible probability $p$, the user needs to consider the sampling time $T$ which is computed within one simulation tick. Initially we defined $T = \SI{0.25}{\second}$, hence a probability $p = \SI{100}{\%}$ implies to generate a maximum number of four vehicles within a simulated period of $\SI{1}{\second}$.
		
		\subsection{Path Computation}
			In order to guide vehicles through a pre-designed map, each traffic component is converted into a graph theory-based representation. For simplicity's sake, we defined each road to consist of two lanes, one for each direction. Furthermore, yatSim focuses on right-handed traffic only\footnote{A transformation into left-handed traffic would be also possible, since the movement directions only project in the opponent direction.}.
			
			Intersections consist of four lanes. One for each possible crossing or turning point. In order to build a valid graph representation, each lane consists of a description of the allowed directions, represented as symbols, shown in Table \ref{tab:symbols}. Lanes, which allow multiple directions, combine these symbols with horizontal direction descriptions as first item and vertical ones as least item.
			
			Within the graph representation each lane is represented by a single vertex. At the simulation's initialization, yatSim iterates through all nodes and connects them with edges according to the directions described in the corresponding lane instances. Whenever a generator introduces a new vehicle into the system, it endows it with a pre-defined random path governed by Dijkstra's algorithm \cite{Dijkstra.1959}, which was performed before on the previously initialized lane graph. Configurability is assured by path filtering, i.e. the user may predefine directions a vehicle is not allowed to turn to, e.g. left-turnings.
			
			\begin{table}
				\caption{Symbols to indicate lane directions in yatSim}
				\label{tab:symbols}
				\begin{tabular}{c c c c}
					\hline \\
					\textbf{Left-To-Right} & \textbf{Right-To-Left} & \textbf{Top-To-Bottom} & \textbf{Bottom-To-Top} \\
					&&&\\
					\hline
					&&&\\
					{$->$} & {$<-$} & {$^\ast\_$} & {$\_^\ast$} \\
					&&&\\
					\hline
				\end{tabular}
				\vspace{-13pt}
			\end{table}
			
			As \cite{Molinari.2018} states, each vehicle $i \in S$ consists of two maps, a local and a global one. A local map function $\mathcal{M}_i$ can transform a local coordinate $p_i(k) \in P_i$ of time instant $k \in \mathbb{N}$ into a global one $\left(x, y\right) \in \mathbb{R}^2$, where $P_i$ indicates the set of all possible local coordinates $i$ may obtain during its ride. An inverse function $\mathcal{M}_i^{-1}$ coherently exists in order to convert global coordinates - if they lay on $i$'s path - into local ones. \cite{Molinari.2018} further proposes to use longitudinal changes on $i$'s path as local coordinates, which are hence applied in the yatSim application as map function $\mathcal{M}_i$.
			
			Whenever a thread starts with the simulation of its corresponding vehicle, the agent checks if it aims to pass an intersection within its residual path and if so, starts to communicate its ambitions to enter it with all competing vehicles, that aim to enter it, too. Based on the result of the hereby performed CBAA-M algorithm, its on-board MPC computes the required acceleration the vehicle requires in order to assure a collision-free passage with the competing agents and a possible frontal vehicle. According to the underlying dynamics model (cf. Equation \ref{eq:vehicle-dynamics:1} and \ref{eq:vehicle-dynamics:2}), the next local position and orientation based on the MPC output is computable. Next, the simulation validates the gathered information on its feasibility, i.e. if the new position is still part of the system. In this case, the vehicle would proceed to move, otherwise it is removed from the simulation, since it reached the end of its entire path.
		
			\begin{figure*}
				\centering
				\subfigure[$i_3$ is current frontal; $i_2$ and $i_1$ have higher priority]{
					\label{fig:shadow:1}
					\resizebox{0.3\textwidth}{0.3\textwidth}{
						\includegraphics{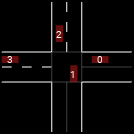}
					}
				}
				~
				\subfigure[$i_1$ is current frontal]{
					\label{fig:shadow:2}
					\resizebox{0.3\textwidth}{0.3\textwidth}{
						\includegraphics{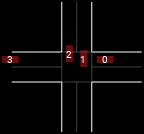}
					}
				}
				~
				\subfigure[$i_2$ is current frontal]{
					\label{fig:shadow:3}
					\resizebox{0.3\textwidth}{0.3\textwidth}{
						\includegraphics{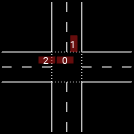}
					}
				}
				\caption{A sample situation of frontal vehicle shadowing out of vehicle $i_0$'s perspective}
				\label{fig:shadowing}
				\vspace{-13pt}
			\end{figure*}
		
		\subsection{Frontal Detection}
			To avoid rear-end collision, each agent in the CBAA-M is responsible to consider the behaviour of its current frontal vehicle and all vehicles, which become a frontal vehicle in the future, e.g. due to sudden turnings into its path. A real-world approach would base on broadcasting the own global position to all vehicles within the system or to all vehicles which are near the current position of a certain radius. In order to reduce the computational simulation efforts, yatSim is aware of all vehicles moving on a dedicated lane, hence it is able to deliver each vehicle its frontal instance by looking up the positions on their paths. In order to gain future frontal vehicles, the MPC assumes each vehicle to stay at constant speed and looks up - based on the agents' individual inverse mapping functions $\mathcal{M}_i^{-1}$ - if the current agent will have passed this point before a certain time or if it still needs to pass it. In the latter case, the other vehicle would be then a frontal candidate in the future.
		
		\subsection{Shadowing Effects}
			During the implementation, we discovered an issue, which we call `shadowing'. Assuming a scenario -  without loss of generality - as in Figures \ref{fig:shadow:1} - \ref{fig:shadow:3}. In this case $i_0$
			approaches the intersection and considers $i_3$ as current frontal vehicle. $i_1$ and $i_2$ are vehicles of higher priority. Since $i_2$ aims to turn right, $i_0$ already identifies $i_2$ as a future frontal vehicle. Hence, $i_0$ adapts its speed in order to avoid a collision with $i_1$ and a possible future collision with $i_2$.
			Within the next simulation `tick', the MPC calculation reruns. Vehicle $i_0$ identifies $i_1$ as current frontal obstacle and further brakes down, in order to avoid a collision. As soon as a vehicle enters the collision point, it stops to bid for it. Hence, $i_1$ and $i_2$ are in this situation not recognized as higher prioritized vehicles to $i_0$. Thus, $i_0$ cannot `see' $i_2$ as a future frontal obstacle. One time step after, $i_1$ has completely passed the collision point and $i_0$ enters the second collision point, which $i_2$ has left before. Since $i_2$ pre-adapted its speed to avoid a collision with $i_3$, the current safety distance between $i_2$ and $i_0$ is too small. Although vehicle $i_0$ now identifies $i_2$ again as the current frontal vehicle, the MPC constraint to keep a minimum safety distance to $i_2$ is infeasible for the next simulated time slot.
			In particular, this scenario implies at least two vehicles to share exactly the same coordinate on their path. In fact, the probability of such a scenario is zero. However, a position quantization, as considered in yatSim, enables this artifact, since coordinates which differed before in the very last decimals are mapped into the same space sample. As \cite{Molinari.2018} does not consider space discretization, it does not need to handle the problem within the theoretical framework. To solve the simulation, we thus configured each vehicle to store the vehicles considered in the previous MPC-computation cycle. In the upper case, this ensures vehicle $i_0$ to remember $i_2$’s future movements, although it is originally shadowed by $i_1$ within the situation of Figure \ref{fig:shadow:2}.
			
			\begin{figure}
				\begin{tikzpicture}
					\node [inner sep=0pt,above right] {\includegraphics[width=\columnwidth]{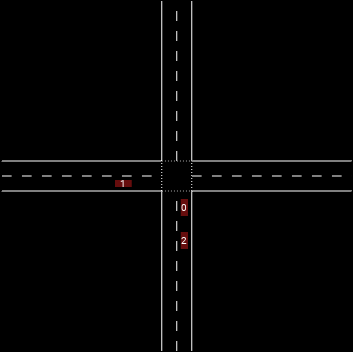}};
					\fill [red] (0.525\columnwidth, 0.48\columnwidth) circle (2pt);
					\node [text=red] at (0.565\columnwidth, 0.475\columnwidth) {\small $h_j$};
				\end{tikzpicture}
				\caption{The simulation's setup: Three autonomous vehicles with different speeds need to find a consensus about the adequate passing order within an intersection of $4\times\SI{40}{\metre}$ long roads}
				\label{fig:sim-setup}
				\vspace{-13pt}
			\end{figure}
		
					\begin{table}
						\caption{Simulation Parameters}
						\label{tab:sim-params}
						\begin{tabular}{cc|p{0.5\columnwidth}}
							\hline &&\\
							\textbf{Parameter} & \textbf{Value} & \textbf{Description} \\
							&&\\
							\hline &&\\
							$H$ & $10$ & The number of horizons the MPC computes with \\
							$L_w$ & $\SI{4.2}{\metre}$ & The length of a single vehicle \\
							$D_w$ & $\SI{1.8}{\metre}$ & The width of a single vehicle \\
							$T_s$ & $\SI{0.25}{\second}$ & The length of a single time sample (i.e. \emph{Sampling Time}) \\
							$p_v$ & $1$ & Weight of speed influence among a vehicle's individual bid (cf. \cite[Equation (15)]{Molinari.2018}) \\
							$p_d$ & $1$ & Weight of distance influence among a vehicle's individual bid (cf. \cite[Equation (15)]{Molinari.2018}) \\
							$\epsilon$ & $0.1$ & Value to avoid numerical errors in vehicle bid equation (cf. \cite[Equation (15)]{Molinari.2018}) \\
							$\lambda_2$ & $\SI{1}{\second}$ & Time required to completely brake a vehicle \\
							$\lambda_3$ & $\frac{L_w}{2} = \SI{2.1}{\metre}$ & Minimum `bumper-to-bumper' distance between two consecutive vehicles \\
							$\underline{v}_i$ & $\SI{0}{\kilo\metre\per\hour}$ & The minimum allowed speed of vehicle $i$ \\
							$\overline{v}_i$ & $\SI{100}{\kilo\metre\per\hour}$ & The maximum allowed speed of vehicle $i$ \\
							$\underline{a}_i$ & $\SI{-9}{\metre\per\square\second}$ & The minimum allowed acceleration of vehicle $i$ \\
							$\overline{a}_i$ & $\SI{5}{\metre\per\square\second}$ & The maximum allowed acceleration of vehicle $i$ \\
							$q$ & $0.1$ & Weight of squared speed difference influence in MPC cost function (cf. \cite[Equation (36)]{Molinari.2018}) \\
							$r$ & $0.01$ & Weight of squared controller valve influence in MPC cost function (cf. \cite[Equation (36)]{Molinari.2018}) \\
							$\omega$ & $0.1$ & Weight of the MPC's soft-constraint variable $\delta$ (cf. \cite[Equation (36)]{Molinari.2018})  \\
							$\mu_\text{ref}$ & $\SI{50}{\kilo\metre\per\hour}$ & Mean vehicle speed to initialize triangular speed distribution \\
							$\sigma_\text{ref}$ & $\SI[parse-numbers=false]{\sqrt{5}}{\kilo\metre\per\hour}$ & Standard deviation to initialize triangular speed distribution
						\end{tabular}
						\vspace{-13pt}
					\end{table}
		
		\subsection{Vehicle Synchronization}
			To assure correct computations, we discovered that all vehicle simulation threads require dedicated synchronization points. For example, situations arise, where a vehicle requires data from another vehicle, which still performs another computation. Therefore, we build up on Python's Threading-Event library a synchronization framework, which stops all threads that completed a previous computation in order to wait on the residual threads before executing the subsequent commands. Those synchronization zones hold all incoming threads, until the last expected thread enters the region. At this point the framework assures the waiting threads to have the same information about the residual ones. When leaving the zone, all threads continue within their process flow as before. The implementation of such a framework becomes crucial, as soon as the vehicle threads completed the CBAA-M and intend to continue with the MPC computation. For a greater number of threads direct communication between the threads becomes impossible, hence they need to wait within the framework until each agent received the final solution.
		
		\subsection{Ghost Threads}
			As a further problem of concurrent simulation, cases occur where particular threads already started the CBAA-M computation and require the result of a sleeping thread. Those issues may arise deadlocks during the simulation. In order to continue the execution flow, we introduced the concept of `ghost' threads. These are duplicates of the original sleeping threads, that are generated during runtime in order to compute the required results. The original thread (`invoker'), which invoked the `ghost' communicates with the `ghost' as if it communicates with the original agent. Hence, it is not able to distinguish, if the current responses are generated from a `ghost' or not. After completion, the `ghost'-thread saves its results into a variable, commonly shared with the original thread. Afterwards, it dies. The `invoker' can thus fluently continue its computations and the sleeping thread already gathered information before it woke up. Nevertheless, as soon as the sleeping thread gets alive and performs possible residual computations, it needs to check, if `ghost' replicates were invoked and are still alive. In this case it needs to wait for them, before proceeding within the execution flow in order to assure a valid computation for each participant.

	\begin{figure}
		\centering
		\includegraphics[width=\columnwidth]{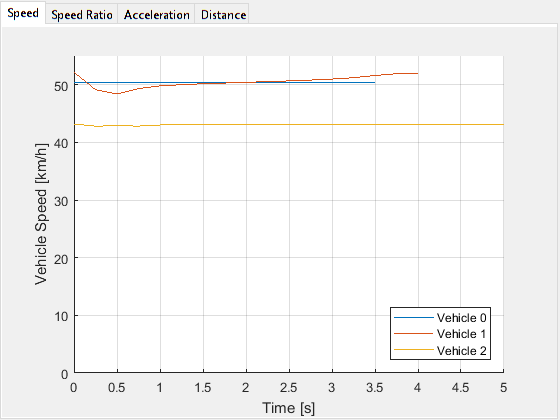}
		\caption{Time behaviour of the vehicles' speed during the simulation run}
		\label{fig:sim:speed}
	\end{figure}
	
	\begin{figure}
		\centering
		\includegraphics[width=\columnwidth]{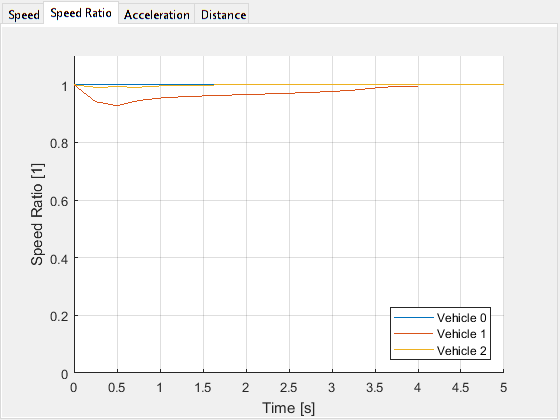}
		\caption{Time behaviour of the vehicles' speed ratio during the simulation run}
		\label{fig:sim:ratio}
		\vspace{-13pt}
	\end{figure}
	
	\section{Simulation}\label{sec:simulation}
		In order to identify the microscopic simulation behaviour, we assumed a similar scenario as in \cite[Figure 3]{Molinari.2018}, shown in Figure \ref{fig:sim-setup}. As total intersection size, we used a road space of $4 \times \SI{40}{\metre}$. Hereby, vehicle $i_0$ starts with a distance of $d_{{i_0}j} = \SI{6}{\metre}$ from the common collision
		point $h_j$. Analogously, vehicle $i_1$ starts with a distance of $d_{{i_1}j} = \SI{11.5}{\metre}$ and vehicle $i_2$ with a distance of $d_{{i_2}j} = \SI{14.25}{\metre}$.
		
		Attentive readers may note that in the reference example of \cite{Molinari.2018}, $d_{{i_2}j}$ was set to $\SI{14}{\metre}$. This difference results from a different parameter setting, listed in Table \ref{tab:sim-params}. As initial speed, $i_0$ drives with a reference speed of $v_{r_{i_0}} = \SI{51}{\kilo\metre\per\hour}$ and intends to turn right, whereas $i_1$ drives with $v_{r_{i_1}} = \SI{53}{\kilo\metre\per\hour}$ and $i_2$ with $v_{r_{i_2}} = \SI{44}{\kilo\metre\per\hour}$. Both, $i_1$ and $i_2$, aim to move forward without any turn. As MPC horizon length $H = 10$ was used, with a sampling time of $T_s = \SI{0.25}{\second}$. Hence, the vehicles cover a frontal distance of around $\SI{30}{\metre}$, which should be more than sufficient for our purpose. We further assumed the vehicles to be connected within a fully-connected network.
		
\begin{figure}
			\centering
			\includegraphics[width=\columnwidth]{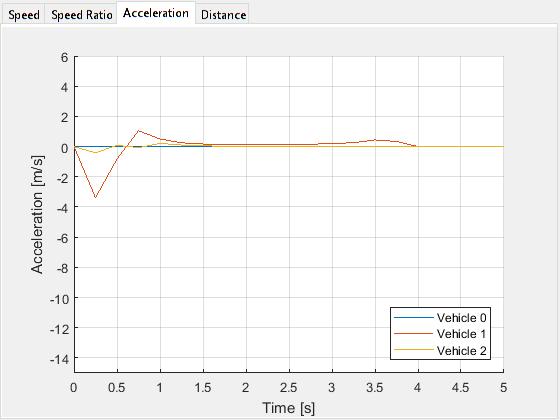}
			\caption{Time behaviour of the vehicles' acceleration during the simulation run}
			\label{fig:sim:acc}
		\end{figure}

		To evaluate the scenario, we measured the vehicles' speed $v_i$ and speed ratio $\frac{v_i}{v_{r_i}}$ (i.e. the ratio of actual speed and desired speed), since CBAA-M aims to enable fast intersection passage and fast transportation with high throughput and low delays; acceleration $a_i$ to investigate the impact of noise, particulate matter distribution, fuel consumption and pollution; and finally the collision point distance $d_{ij}$ for $h_j$ to investigate how fast a vehicle frees the point for residual agents. During the simulation the vehicles determined the passing order of $i_0 \rightarrow i_1 \rightarrow i_2$. Considering the acceleration and speed information (cf. Figures \ref{fig:sim:acc} and \ref{fig:sim:speed}), vehicle $i_1$ broke in order to let vehicle $i_0$ pass. Afterwards it accelerates fast in order to pass the intersection before vehicle $i_2$. Both, agent $i_0$ and $i_2$, can hold during the complete simulation a speed ratio of approx. $\SI{100}{\%}$, whereas $i_1$ needs to consider its distance to $i_0$ as frontal vehicle, which is slower than $i_1$. Hence $i_1$'s speed ratio is decreased until $i_0$ leaves the simulated scenario.
	
		\begin{figure}
			\centering
			\includegraphics[width=\columnwidth]{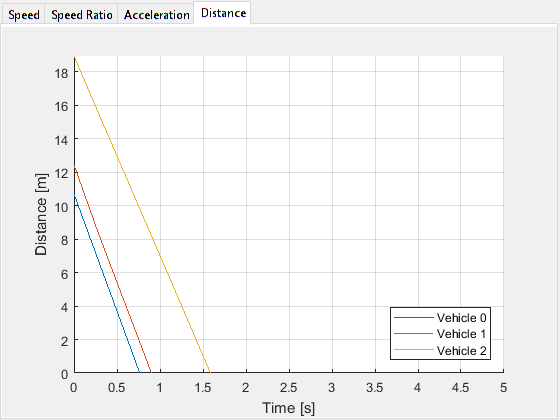}
			\caption{Time behaviour of the vehicles' collision point distance during the simulation run}
			\label{fig:sim:dist}
			\vspace{-13pt}
		\end{figure}
	
	\section{Conclusion}\label{sec:conclusion}
	With yatSim we provide a suitable and dynamic framework to implement and measure the behaviour of autonomous vehicles within urban traffic scenarios. Starting at the application description, we presented our reasons for the selection of Python and the Kivy framework as implementation tools. We further showed the simulation process based on a double-clocking approach. Thereon, we introduced the Monte-Carlo vehicle generation process and gave an insight on the Dijkstra path computation. Considering the implementation of the CBAA-M, we outlined the mapping approach to discover frontal vehicles. By that, we discovered and solved the `shadowing' issue. Finally, we considered the thread synchronization and `ghost threads' in order to enable multi-threaded simulation.
	
	We concluded the present work with a simple microscopic traffic simulation, which promotes the ability of yatSim to simulate CBAA-M in an urban scenario. In further tests, which are not presented in the context of this paper, we also investigated the impact on macroscopic simulations, where multiple inter-connected road nets with more than thousand interacting vehicles were simulated.

	However, as any other software product, also this application benefits from future developments. Although the development focused to strictly encapsulate code, errors may arise during code change. Additional unit tests and similar code quality assurance features may reduce this impact. For the user-side, several features might increase the experience, as e.g. an internal application history to redo or undo actions within the designer environment. Also the	auto-routing might perceive changes, as e.g. a preceding path designer module, which 	 enables the user to further influence vehicle behaviours for detailed studies. Another future topic will be the simulation of network issues in the vehicle-to-vehicle communications to investigate CBAA-M's convergence-robustness and stability under non-fully-connected constraints.
	
	As final product, we published this software on GitLab\footnote{https://gitlab.com/fgrs/yatsim} with open access to everyone who is interested to take part in the development or to use it as a measuring tool.

\bibliography{literature}
\end{document}